\documentclass[12pt]{iopart}
\usepackage{graphicx}

\begin{document}

\title{First-order phase transition in a 2D random-field Ising model with conflicting dynamics}
\author{Nuno Crokidakis}
\address{
Instituto de F\'{\i}sica - Universidade Federal Fluminense \\
Av. Litor\^anea, s/n \\
24210-340 \hspace{5mm} Niter\'oi - Rio de Janeiro \hspace{5mm} Brazil}

\ead{nuno@if.uff.br}

\begin{abstract}

The effects of locally random magnetic fields are considered in a nonequilibrium Ising model defined on a square lattice with nearest-neighbors interactions. In order to generate the random magnetic fields, we have considered random variables $\{h\}$ that change randomly with time according to a double-gaussian probability distribution, which consists of two single gaussian distributions, centered at $+h_{o}$ and $-h_{o}$, with the same width $\sigma$. This distribution is very general, and can recover in appropriate limits the bimodal distribution ($\sigma\to 0$) and the single gaussian one ($ho=0$). We performed Monte Carlo simulations in lattices with linear sizes in the range $L=32 - 512$. The system exhibits ferromagnetic and paramagnetic steady states. Our results suggest the occurence of first-order phase transitions between the above-mentioned phases at low temperatures and large random-field intensities $h_{o}$, for some small values of the width $\sigma$. By means of finite size scaling, we estimate the critical exponents in the low-field region, where we have continuous phase transitions. In addition, we show a sketch of the phase diagram of the model for some values of $\sigma$.
\end{abstract}


\maketitle

\section{Introduction}

The Random Field Ising Model (RFIM) is one of the most studied systems in magnetism (for reviews, see \cite{binder_review,belanger_review} and more recently \cite{dotsenko}). Its simple theoretical definition and the interesting physical properties that emerge from its study represent two great motivations for the investigation of this model. In addition, a considerable experimental interest arised after the identification of the RFIM with some diluted antiferromagnets in the presence of a uniform magnetic field \cite{fishman,cardy}, like ${\rm Fe_{x}Zn_{1-x}F_{2}}$ and ${\rm Fe_{x}Mg_{1-x}Cl_{2}}$ \cite{belanger_review,belanger,birgeneau}.

Following the success of the Ising model to capture the essential physics of complex systems, several lattice models have been proposed \cite{dickman}. These models involve a periodic lattice whose sities are occupied by spin-$1/2$ variables with two states, $s_{i}=\pm 1$. Every configuration, $\textbf{s}=\{s_{i}\}$, has a potential energy given by the Ising Hamiltonian,
\begin{equation}\label{eq1}
\mathcal{H}(\textbf{s})=- \sum_{<i,j>}J_{ij}s_{i}s_{j} - \sum_{i}h_{i}s_{i}~, 
\end{equation}
\noindent
i.e., any pair of nearest-neighbors spins contributes with a random exchange interaction $J_{ij}$ and each spin interacts also with a magnetic field of local intensity $h_{i}$. Systems based on Eq. (\ref{eq1}) are called models of \textit{quenched} disorder (MQD) \cite{dickman}. They are characterized by a frozen-in spatial distribution of disorder, i.e., $J_{ij}$ and/or $h_{i}$ vary at random with $i$ but remains fixed with time. These models have been very well explored in the literature \cite{belanger_review}, and some experimental results can theoretically be reproduced even at mean-field level \cite{meu}. However, MQD are \textit{equilibrium} models, and they neglect diffusion of magnetically active ions that occurs in some systems, like spin glasses \cite{belanger_review}. Diffusion constantly modifies the distance between each specific pair of spin ions in substances and, consequently, one should probably allow for variations both in space and time of $J_{ij}$ and/or $h_{i}$ in a model. These effects do not seem to be correctly described by \textit{annealed} systems, where the change with time of the spatial distribution of $J_{ij}$ and/or $h_{i}$ is constrained by the need to reach equilibrium with the other degrees of freedom \cite{thorpe,urumov,haroni,goncalves,vieira}. Therefore, impurities tend to be strongly correlated, which is not observed in most substances. Instead, one may conceive a situation in which both $\textbf{s}$ and the spatial distribution of impurities $J_{ij}$, or the spatial distribution of the local fields $h_{i}$, vary with time. That is, one may assume that spins and impurities (or fields) behave more independently of each other than in the annealed model so that a conflict occurs, and a steady \textit{nonequilibrium} condition prevails asymptotically. This is consistent with the reported observation of nonequilibrium effects, for exemple, the influence of the details of the dynamical process on the steady state in some materials \cite{dickman}. These models can be called nonequilibrium random-field Ising models (NRFIM)

The study of nonequilibrium models defined by Eq. (\ref{eq1}) reveals many interesting features \cite{dickman}, with a rich variety of phase transitions and critical phenomena. Exact results obtained for one-dimensional lattices \cite{lacomba1,lacomba2,garrido} and Monte Carlo (MC) simulations \cite{miranda,bonilla} reveal that the critical behavior is non-universal, but it generally depends on apparently irrelevant details of dynamics, like diffusion of impurities, i.e., the properties of the distribution of the random variables, and on the transition rates chosen. The probability distributions analyzed were discrete, for random exchange interactions and/or random magnetic fields \cite{dickman}. As experimental realizations of these systems, we may imagine : i) a magnetic material under the action of a random (or very rapidly fluctuating) magnetic field, i.e., a field that varies according to a given probability distribution with a period shorter than the mean time between successive transitions that modify the spin configuration, or ii) a disordered system with fast and random diffusion of impurities in which the latter consist of both exchanges as in spin glasses and local fields as in random field systems \cite{lacomba1}.

Due to these motivations, we have studied the NRFIM on a square lattice with nearest-neighbors interactions and in the presence of random magnetic fields that follow a double-gaussian probability distribution. We performed Monte Carlo simulations and our results suggest that first-order phase transitions occur in the model for some values of $\sigma>0$ at low temperatures and high random-field intensities. In the low-field region, we have performed a finite size scaling (FSS) to estimate the critical exponents, in order to test whether the system follows continuous FSS laws.


\section{Model and Monte Carlo Simulation}

A simple realization of the ideas discussed in the introduction is the Kinetic Ising Model \cite{dickman}, where the spin system is in contact with a heat bath at temperature $T$ that induces stochastic changes of $\textbf{s}$ according the master equation \cite{miranda}
\begin{equation}\label{eq2}
\frac{\partial{P_{t}(\textbf{s})}}{\partial{t}}=\sum_{\textbf{s}^{i}}[c(\textbf{s}^{i};i)P_{t}(\textbf{s}^{i}) -c(\textbf{s};i)P_{t}(\textbf{s})].
\end{equation}
\noindent
Here, $P_{t}(\textbf{s})$ is the probability of any configuration $\textbf{s}$ at time $t$, and $c(\textbf{s};i)$ is the probability per unit time for a transition from $\textbf{s}$ to $\textbf{s}^{i}$; the latter is obtained from $\textbf{s}$ by flipping spin $s_{i}$. The detailed balance condition, i.e., $c(\textbf{s};i)= c(\textbf{s}^{i};i)\,\exp[-\beta\Delta\mathcal{H}]$, with $\Delta\mathcal{H}=\mathcal{H}(\textbf{s}^{i})-\mathcal{H}(\textbf{s})$, where $\beta=(k\,T)^{-1}$ is the inverse temperature ($k$ is the Boltzmann constant), is sufficient to guarantee that the stationary solution of Eq. (\ref{eq2}) is the Gibbs state corresponding to energy given by Eq. (\ref{eq1}) and temperature $T$ \cite{dickman,miranda}. This is satisfied by the Metropolis algorithm \cite{metropolis}, $c(\textbf{s};\textbf{i})={\rm min}\{1, \exp(-\beta\Delta\mathcal{H})\}$, for instance \cite{miranda}.

We have considered the Hamiltonian given by Eq. (\ref{eq1}) on a square lattice of linear dimension $L$ and constant exchange interaction, i.e., $J_{ij}=J>0$. In other words,
\begin{equation}\label{eq3}
\mathcal{H}=- J\sum_{<i,j>}s_{i}s_{j} - \sum_{i}h_{i}s_{i}, 
\end{equation}
\noindent
and the random fields $\{h_{i}\}$ follow a continuous probability distribution, namely, the double-gaussian one,
\begin{equation}\label{eq4}
P(h_{i})=\frac{1}{2}\frac{1}{\sqrt{2\pi\sigma^{2}}}\left\{\exp\left[-\frac{(h_{i}-h_{o})^{2}}{2\sigma^{2}}\right] + \exp\left[-\frac{(h_{i}+h_{o})^{2}}{2\sigma^{2}}\right]\right\},
\end{equation}
\noindent
where we have two gaussian distributions centered at $\pm h_{o}$ with the same width $\sigma$. The time evolution of any state $\textbf{s}$ is generated by Eq. (\ref{eq2}) with $c(\textbf{s};\textbf{x})$ describing a competing process \cite{miranda}. For the numerical implementation, the algorithm is as follows: at each time step, a new configuration of random fields $\{h_{i}\}$ is generated according to $P(h_{i})$, Eq. (\ref{eq4}); then, every lattice site is visited, and a spin flip occurs according to Metropolis' rule. In other words, the random variables $h_{i}$ vary with time, i.e., the system is described at each time by Eq. (\ref{eq3}), with $h_{i}$ distributed according to $P(h_{i})$ given by Eq. (\ref{eq4}). The \textit{instantaneous} Hamiltonian, Eq. (\ref{eq3}), corresponds to the one that characterizes the quenched RFIM. However, the local fields also change at random with time at each site according to $P(h_{i})$. Consequently, a sort of dynamic conflict occurs, which differs fundamentally from the \textit{equilibrium} quenched and annealed random-field models, and we have a \textit{nonequilibrium} system \cite{dickman}. We may interpret this as a magnet under a field that is changing (all over system) at random with time according to $P(h_{i})$ \cite{lacomba1,lacomba2}. Thus, we have two different characteristic time scales: one for the fluctuations of the spins and another one for the fluctuations of the random field, and we consider that these two fluctuations are independent (a formal discussion about this is found in ref. \cite{dickman}, chapter 7). We will show that this rapid fluctuation of the random variables $h_{i}$ during the time evolution of the system lead to a first-order transition between the ordered and the disordered phases for high values of the random-field intensity $h_{o}$, in opposition of the continuous transition that was found for small $h_{o}$.

\begin{figure}[t]
\begin{center}
\vspace{1.0cm}
\includegraphics[width=0.4\textwidth,angle=0]{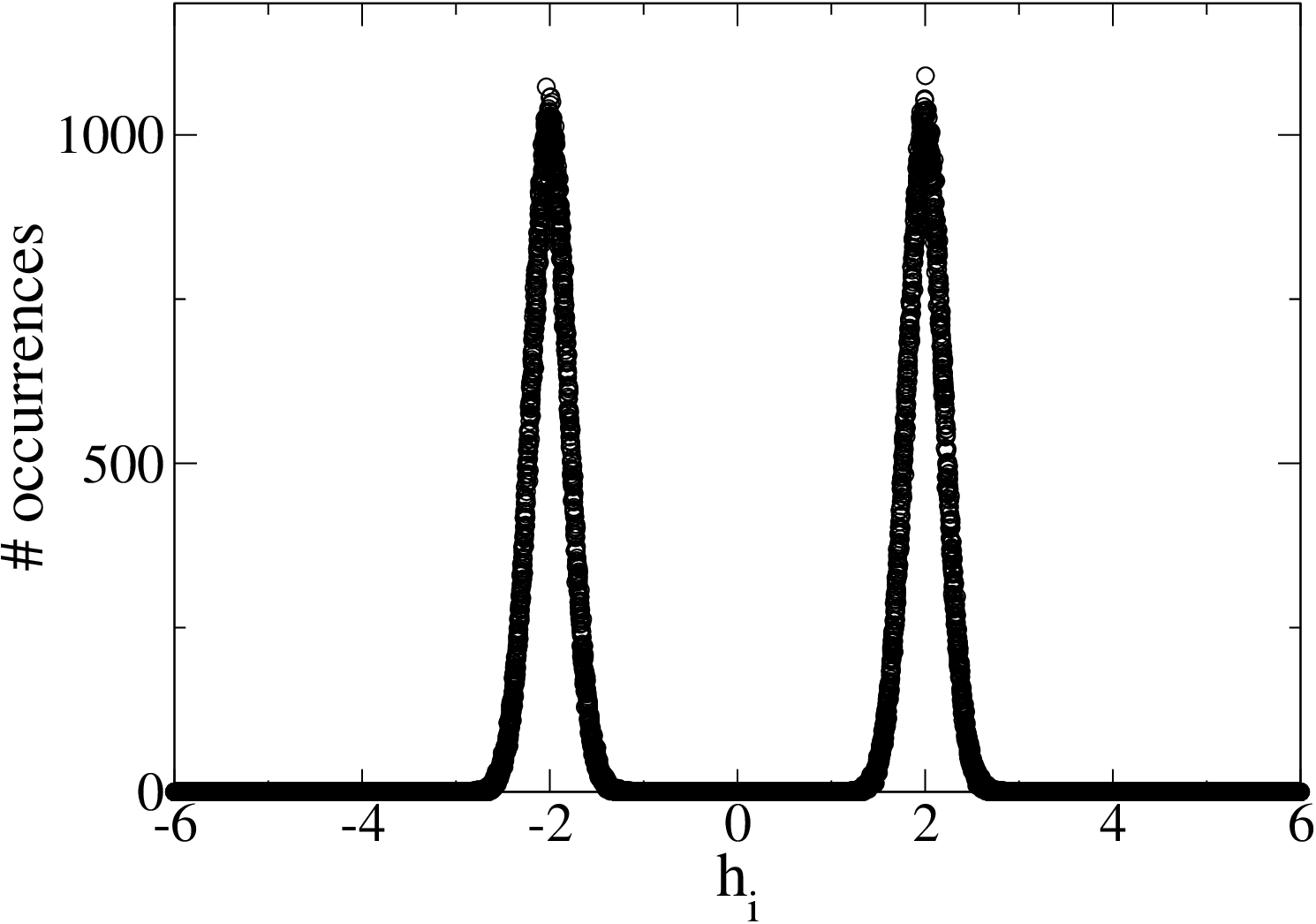}
\hspace{0.5cm}
\includegraphics[width=0.4\textwidth,angle=0]{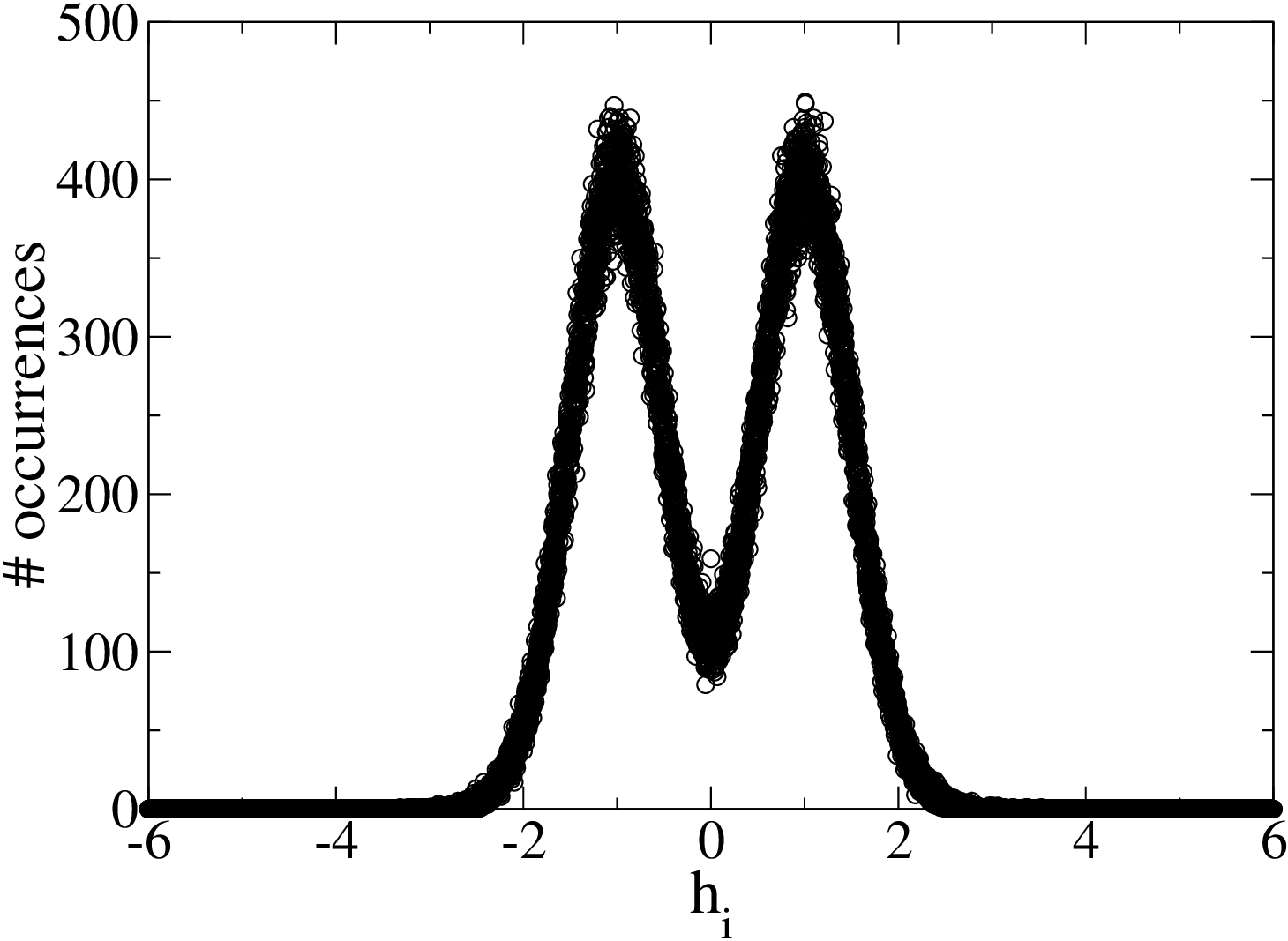}
\end{center}
\caption{Histograms of the random variables $\{h_{i}\}$ generated numerically accordingly Eq. (\ref{eq4}), for $\sigma=0.2$ and $h_{o}=2.0$ (left) and $\sigma=0.5$ and $h_{o}=1.0$ (right).}
\label{Fig1}
\end{figure}

Although we do not have evidences that the double-gaussian distribution for the random fields can be realized experimentally, we can argue that this distribution is suitable for an appropriate theoretical description of random-field models \cite{meu}. In the identification of the RFIM with diluted antiferromagnets in the presence of a uniform magnetic field, the local random fields are expressed in terms of quantities that vary in both signal and magnitude \cite{fishman,cardy}. This characteristic rules out the bimodal probability distribution from a such class of physical systems. Although the RFIM defined in terms of a single gaussian probability distribution is physically acceptable, it usually leads only to continuous phase transitions, either within mean-field \cite{aharony,schneider,andelman}, or standard short-range-interaction approaches \cite{gofman,swift}. This discussion is based on quenched random-field models, but an extension to the case of nonequilibrium diluted antiferromagnets was discussed by some authors \cite{dickman,miranda}. As pointed in a previous paper \cite{meu}, the double-gaussian distribution can recover in appropriate limits the bimodal distribution ($\sigma\to 0$) and the single gaussian one ($h_{o}=0$ or $\sigma>h_{o}$). In Fig.\;\ref{Fig1}, we show two histograms of the random variables $\{h_{i}\}$ generated numerically accordingly the probability distribution of Eq. (\ref{eq4}).

\begin{figure}[t]
\begin{center}
\vspace{1.0cm}
\includegraphics[width=0.4\textwidth,angle=0]{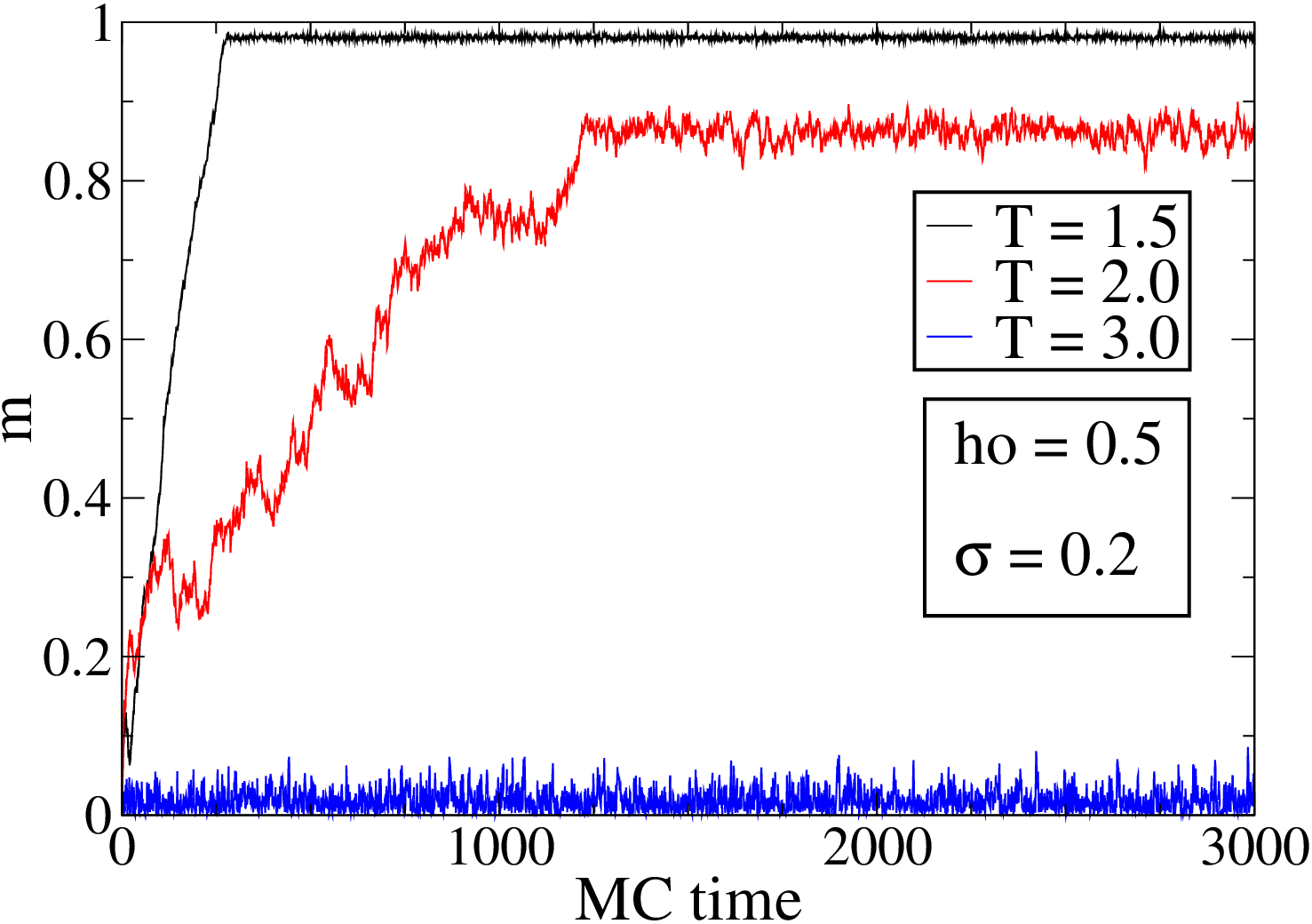}
\hspace{0.5cm}
\includegraphics[width=0.4\textwidth,angle=0]{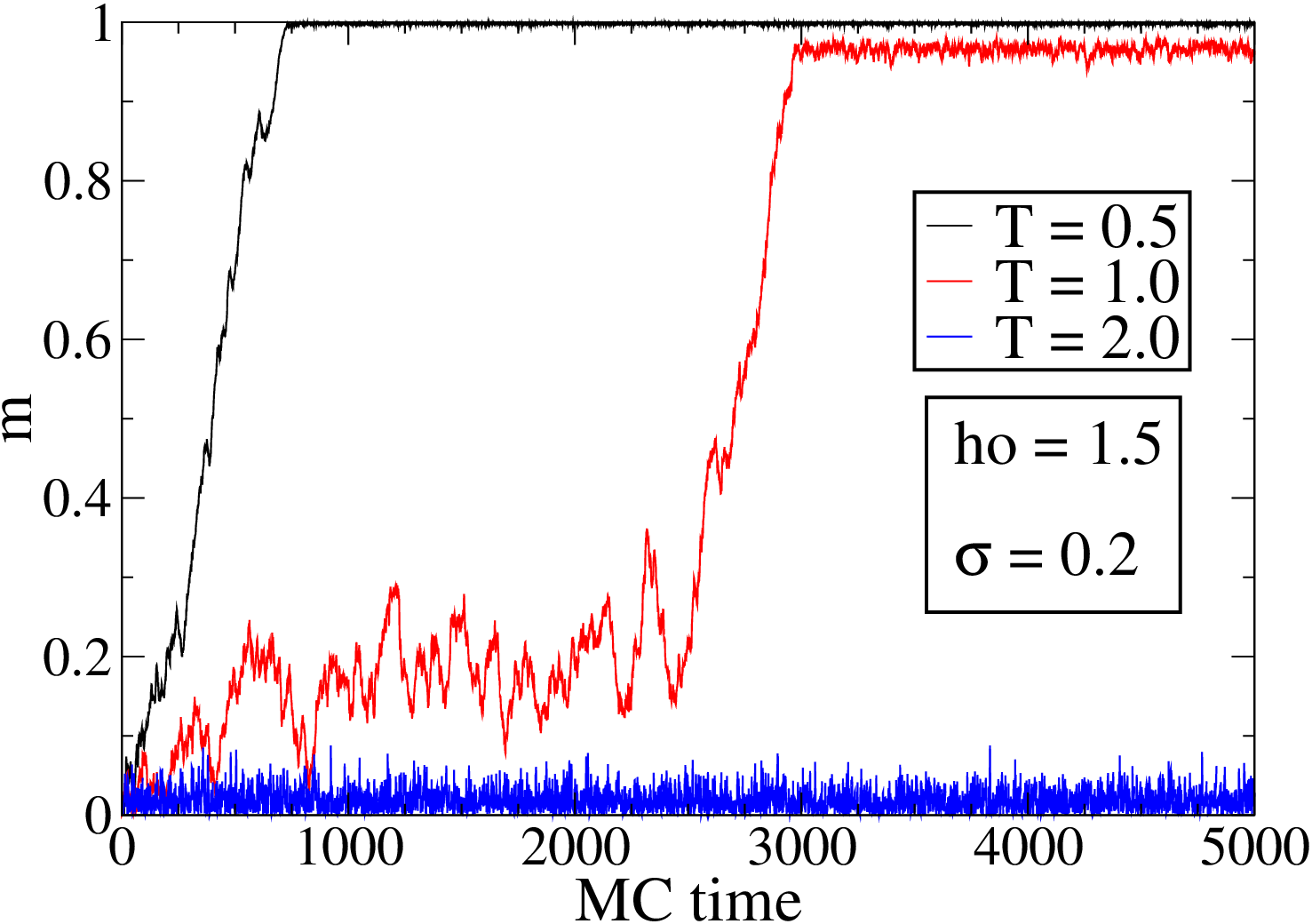}
\end{center}
\caption{Evolution of the magnetization as a function of the Monte Carlo time, for $L=128$ and $\sigma=0.2$. Examples for low (left side) and high random field intensity $h_{o}$ (right side) are shown. As discussed in the text, the equilibrium states are easily anchieved for low $h_{o}$ but are harder to obtain as $h_{o}$ increases.}
\label{Fig2}
\end{figure}

In the following we use for simplicity $J=1$. We have studied systems of $L=32, 64, 128, 256$ and $512$ with periodic boundary conditions and a random initial configuration of the spins. We have analyzed the following values of the parameters: $\sigma=0.0, 0.1, 0.2, 0.3, 0.4$ and $0.5$, all of them in the range $0.0<ho<3.0$ and $0.01<T<4.0$. The results for all values of the magnetic field parameters $\sigma$ and $h_{o}$ show that finite-size effects are less-pronounced for $L\geq 128$. We can test the equilibration of the system by monitoring the magnetization as a function of the MC time. We have found that the equilibrium is easily achieved for low $h_{o}$ but is harder to obtain as the intensity of the random field increases [see Fig.\;\ref{Fig2}]. Thus, we have used $10^{4}$ MC steps for equilibration and $10^{6}$ MC steps for averaging. In addition, we have to consider the autocorrelation function, defined as \cite{gould}
\begin{eqnarray}\label{eq_new}
C(t)=\frac{<m(t+t_{o})m(t_{o})>-<m>^{2}}{<m^{2}>-<m>^{2}} ,
\end{eqnarray}
\noindent
where $m(t)$ is the value of the magnetization of the system at time $t$. The averages in Eq. (\ref{eq_new}) are over all possible time origins $t_{o}$ for a steady state. For sufficiently large $t$, $m(t)$ and $m(0)$ will become uncorrelated, and hence $<m(t+t_{o})m(t_{o})> \to <m(t+t_{o})><m(t_{o})>=<m>^{2}$, i.e., $C(t)\to 0$. In the above-described dynamics, the autocorrelation function $C(t)$ decays in the traditional exponential form, $C(t)\sim e^{-t/\tau}$, where $\tau$ is the autocorrelation time, whose magnitude depends on the choice of the physical parameters of the system ($T$, $h_{o}$, $\sigma$). Thus, to obtain configurations that are statistically independent, we have used $30$ MC steps between two measurements of the quantities of interest (remember that in the standard 2D Ising model, 10 MC steps are sufficient).

\section{Results}

In the following we will show results for the order parameter $m$, the magnetization per spin, and for $\chi$ , the magnetic susceptibility, which can be obtained from the simulations by the fluctuation-dissipation relation,
\begin{equation}\label{eq_chi}
\chi=\frac{<m^{2}>-<m>^{2}}{kT},
\end{equation}
where $k$ is the Boltzmann constant and $<\;>$ stands for MC average \cite{miranda}. For all following results we use $k=1$.

\begin{figure}[t]
\begin{center}
\vspace{1.0cm}
\includegraphics[width=0.5\columnwidth,angle=0]{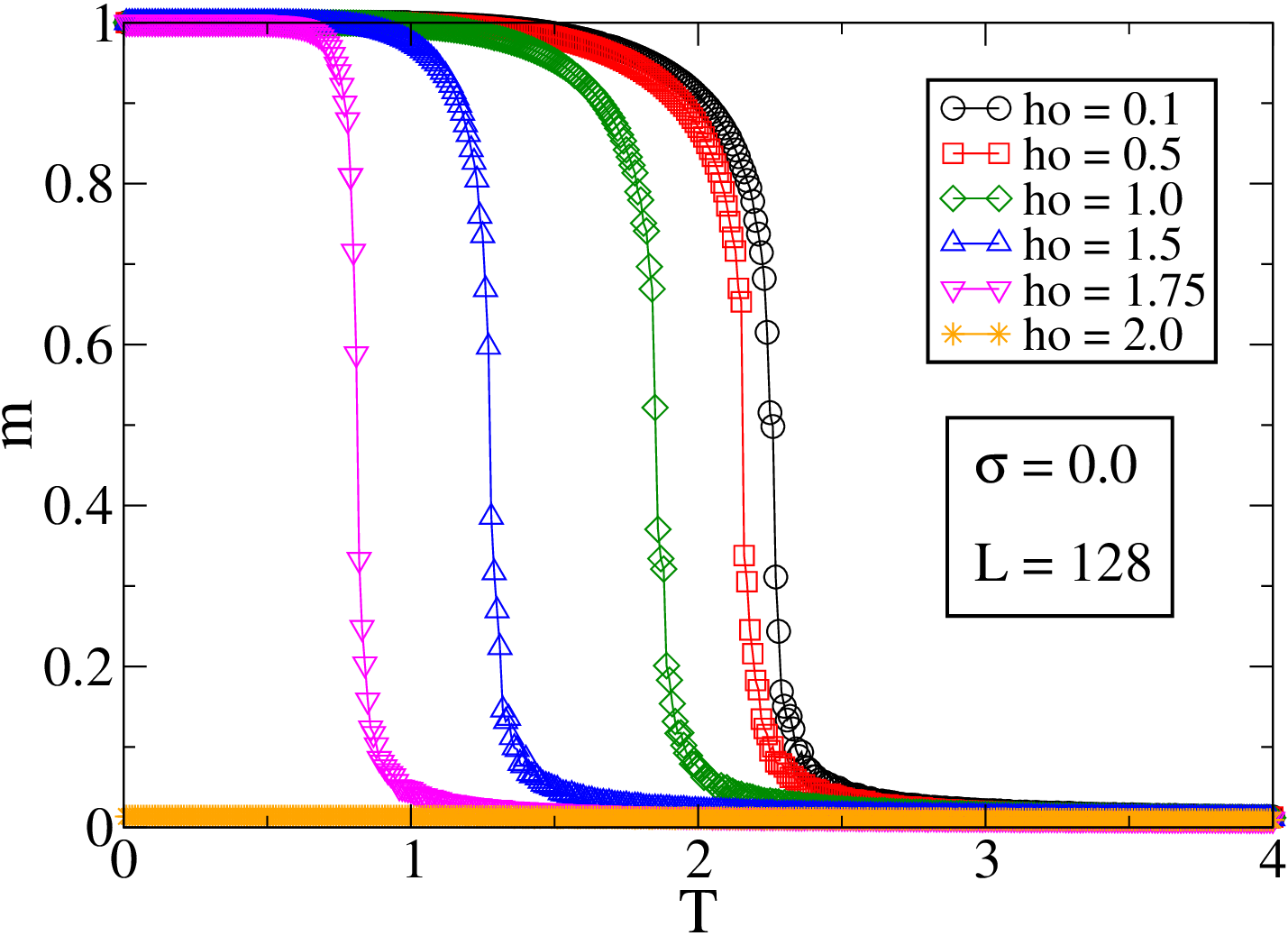}
\end{center}
\caption{Magnetization per spin as a function of temperature for $L=128$, $\sigma=0.0$ and some typical values of the magnetic field intensity $h_{o}$. For high values of $h_{o}$ the system is in the paramagnetic phase, but for all other values of $h_{o}$ we can observe a continuous transition between the ordered and the disordered phases. The symbols (circle, square, diamond, star and triangles up and down) are computed points, whereas the lines are just guides to the eye. The error bars are not shown because they are smaller than data points.}
\label{Fig3}
\end{figure}

\begin{figure}[t]
\begin{center}
\vspace{1.0cm}
\includegraphics[width=0.5\columnwidth,angle=0]{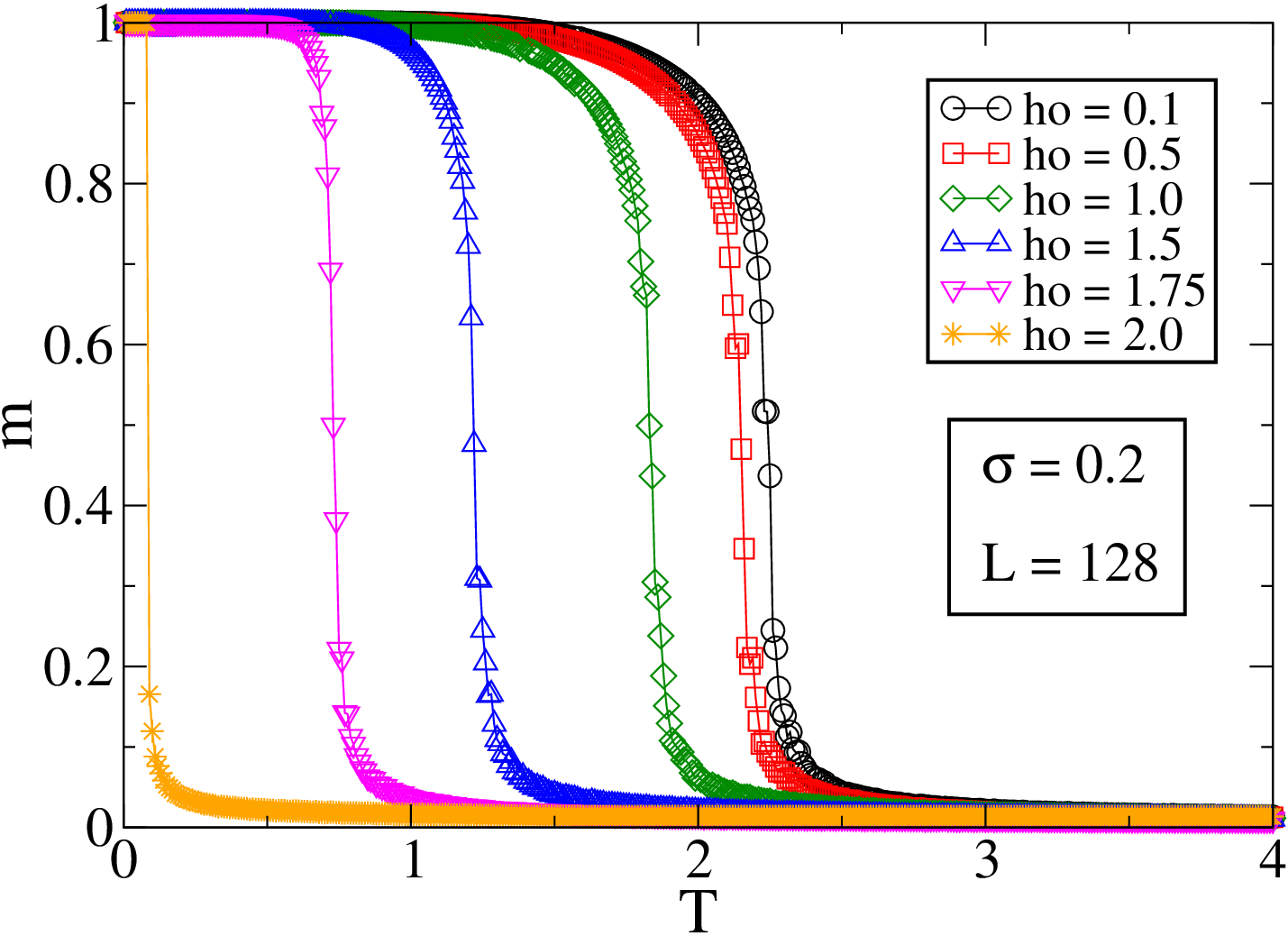}
\end{center}
\caption{Magnetization per spin as a function of temperature for $L=128$, $\sigma=0.2$ and some typical values of the magnetic field intensity $h_{o}$. We can observe that $m$ decreases continuously with $T$ for small values of $h_{o}$, but we can identify jumps on the magnetization for high magnetic-field intensities. The error bars are smaller than data points.}
\label{Fig4}
\end{figure}

Our first simulations were done for $\sigma=0.0$, the limit case of the bimodal distribution. In Fig.\;\ref{Fig3} we show the results of the magnetization per spin, $m$, versus the temperature, $T$, for $L=128$ and some typical values of $h_{o}$. These results show that the system is in the disordered paramagnetic phase for high values of $h_{o}$ ($\geq 2.0$), but for all values $h_{o}<2.0$ there are continuous phase transitions between the ferromagnetic ($m\neq0$) and the paramagnetic ($m=0$) phases.

\begin{figure}[t]
\begin{center}
\vspace{1.0cm}
\hspace{1.0cm}
\includegraphics[width=0.5\columnwidth,angle=0]{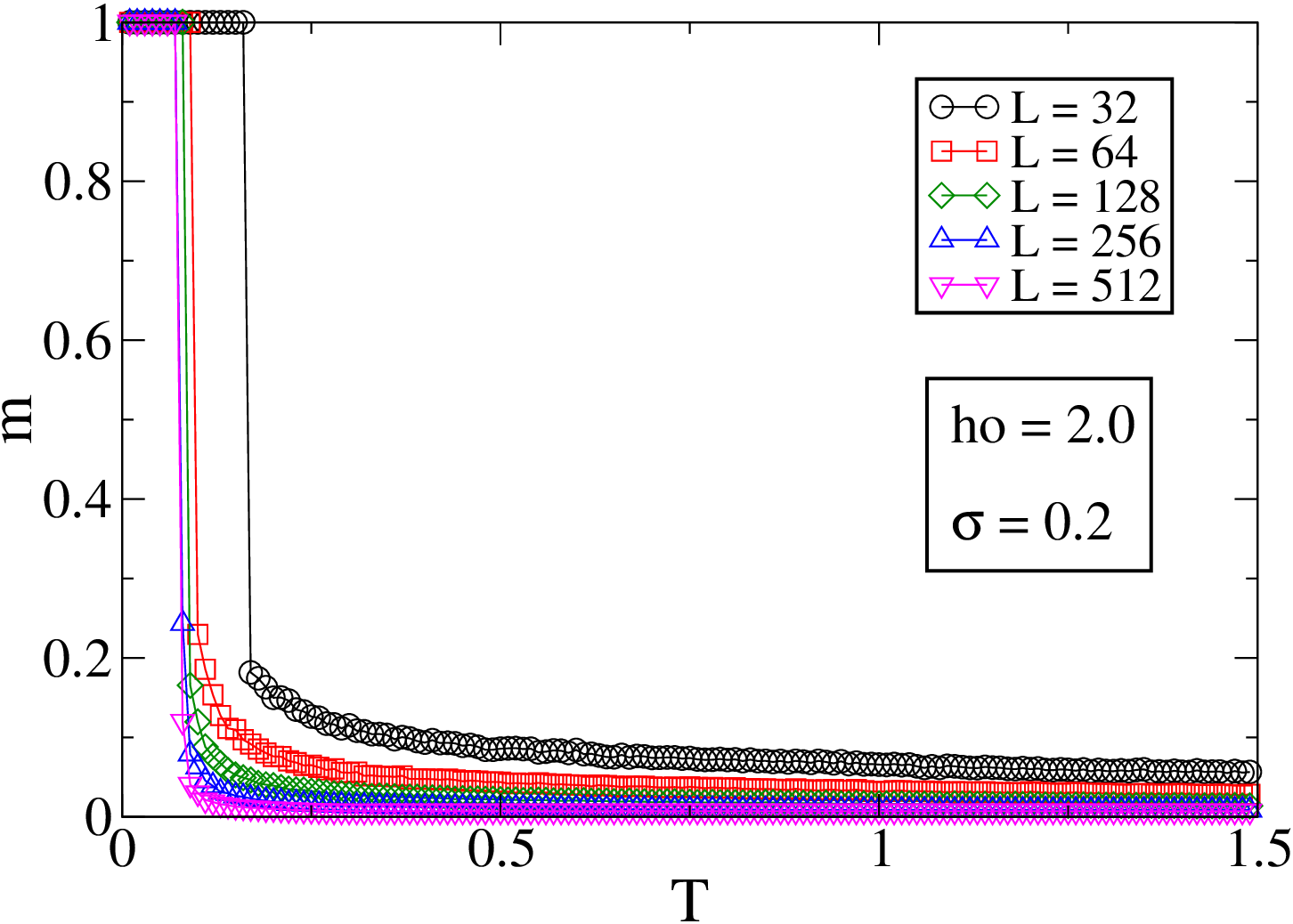}
\end{center}
\caption{Magnetization per spin as a function of temperature for $\sigma=0.2$, $h_{o}=2.0$ and different lattice sizes. We can observe jumps on the magnetization for all lattice sizes. The error bars are smaller than data points.}
\label{Fig5}
\end{figure}

We can analyze now the behavior of the magnetization for $\sigma>0.0$. In Fig.\;\ref{Fig4} we show the magnetization per spin as a function of the temperature for $L=128$, $\sigma=0.2$ and some typical values of $h_{o}$. In this case, for small $h_{o}$ we can observe the same behavior as in the case of $\sigma=0.0$. The difference appears for higher values of the magnetic-field intensity: we can observe jumps in the magnetization for values near $h_{o}=2.0$. So, the next step is to verify the behavior of the system for $\sigma=0.2$, $h_{o}=2.0$ and different linear lattice sizes $L$. In Fig.\;\ref{Fig5} we show the results of magnetization versus the temperature for $L=32, 64, 128, 256$ and $512$. As discussed before, for the smaller value of $L$ we have pronounced finite-size effects. Thus, for the calculations of the exponents, we will discard the size $L=32$.

\begin{figure}[t]
\begin{center}
\vspace{1.0cm}
\includegraphics[width=0.4\columnwidth,angle=0]{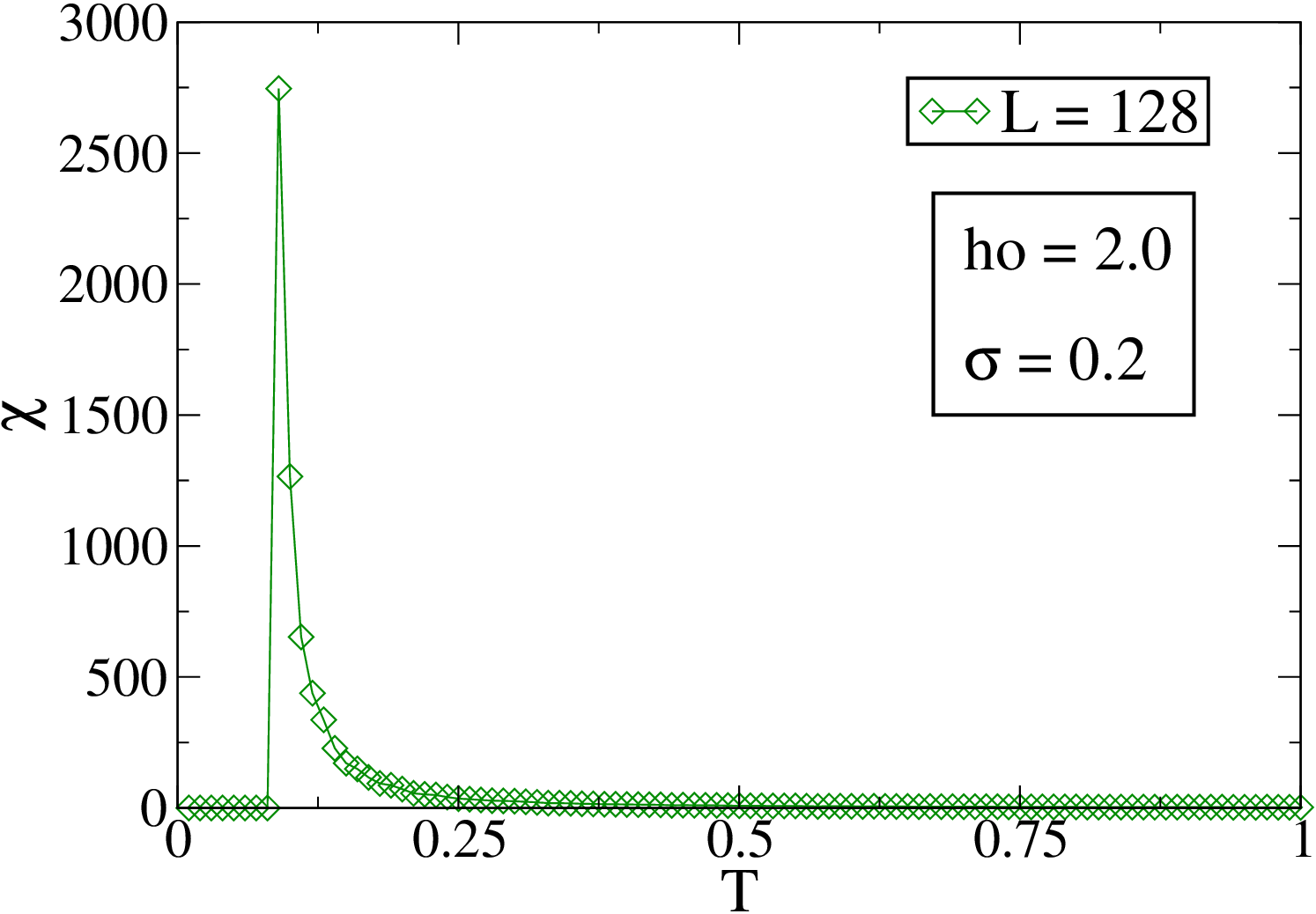}
\hspace{0.5cm}
\includegraphics[width=0.4\columnwidth,angle=0]{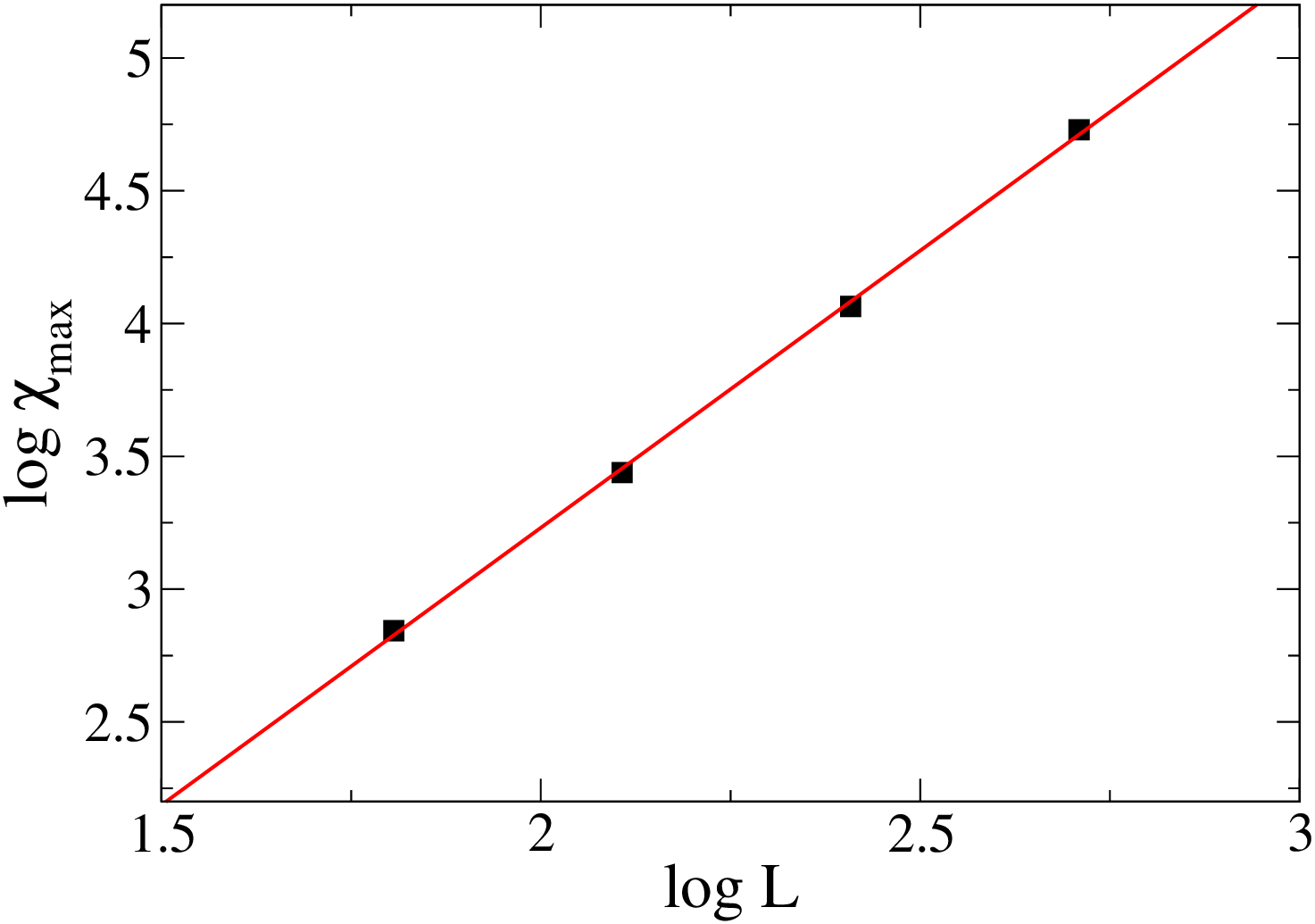}

\vspace{1.0cm}

\includegraphics[width=0.4\columnwidth,angle=0]{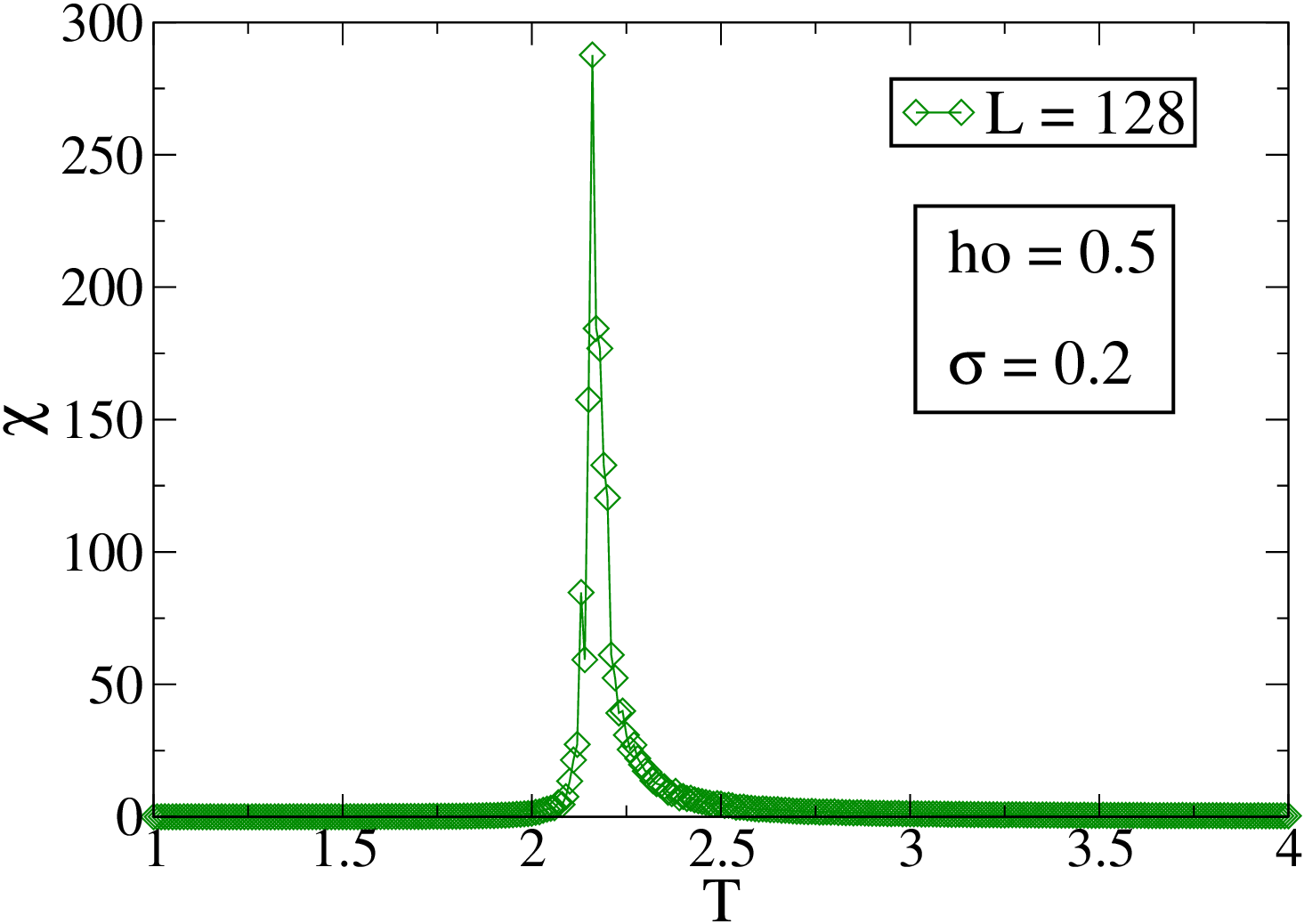}
\hspace{0.5cm}
\includegraphics[width=0.4\columnwidth,angle=0]{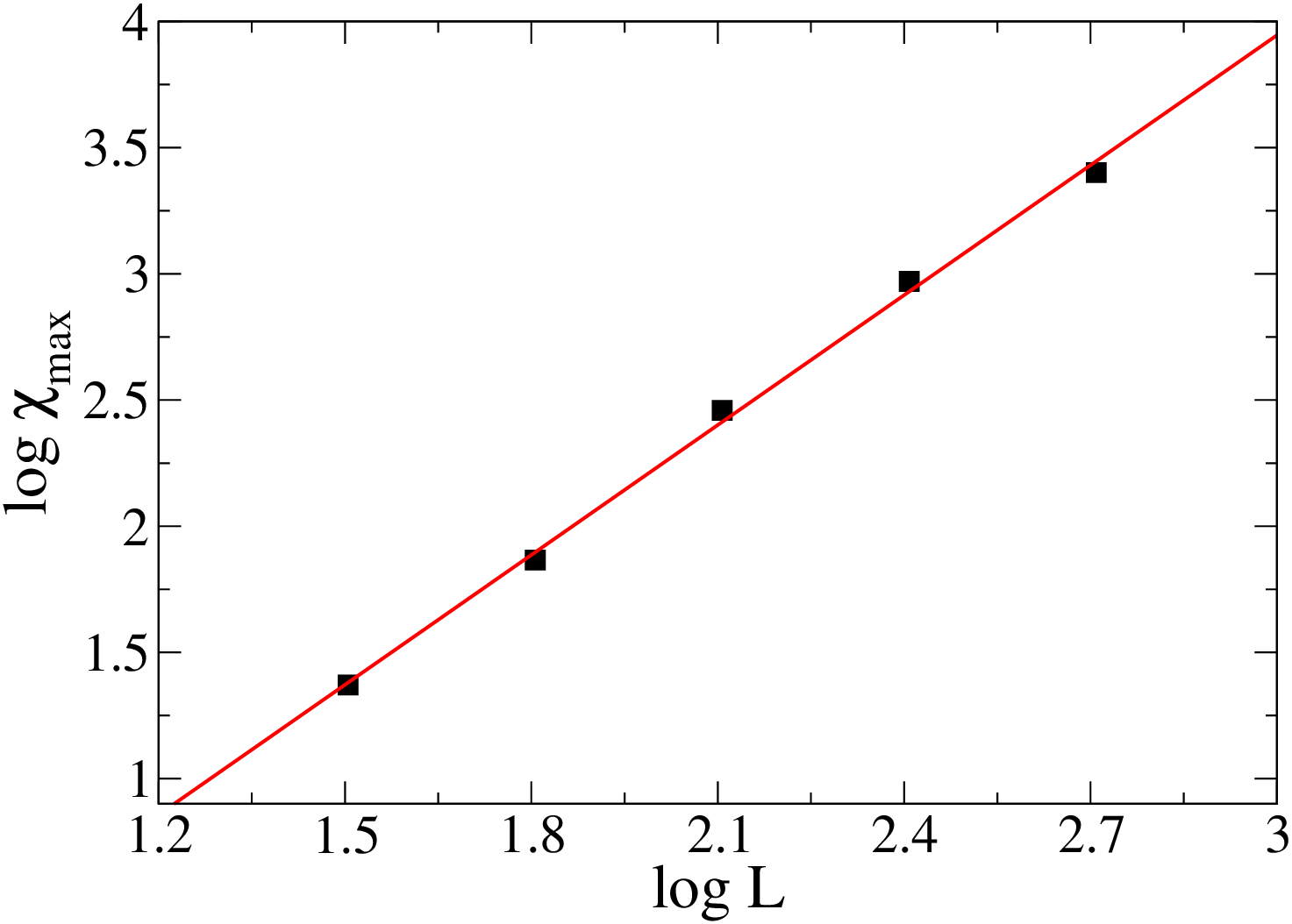}
\end{center}
\caption{In the upper figures we show the susceptibility $\chi$ as a function of the temperature for $\sigma=0.2$, $h_{o}=2.0$ and $L=128$ (left side) and the susceptibility peaks $\chi_{max}$ versus $L$ in the log-log scale for some lattice sizes (right side). In the lower figures we also show the susceptibility $\chi$ as a function of the temperature for $\sigma=0.2$ and $L=128$, but for $h_{o}=0.5$ (left side) and the susceptibility peaks $\chi_{max}$ versus $L$ in the log-log scale for some lattice sizes (right side). The straight lines have slope equal to $2.087$ (for $h_{o}=2.0$) and $1.7157$ (for $h_{o}=0.5$), respectively. As explained in the text, the peaks grow fast and we only show the result for $\chi$ for one lattice size. We have discarded the smaller lattice size $L=32$ for the first-order transition calculation (upper figure, right side), as explained in the text. The error bars are smaller than data points.}
\label{Fig6}
\end{figure}

The jumps observed in the magnetization in Fig.\;\ref{Fig5} suggest a first-order ferromagnetic-paramagnetic phase transition at low temperatures. In order to verify this suggestion, we shall analyze the susceptibility $\chi$, Eq. (\ref{eq_chi}). We have verified that the susceptibility peak grows very fast with the linear lattice size $L$, which rules out the possibility of the plot of $\chi$ versus $T$ for all lattice sizes studied. Due to this, we have plotted in Fig.\;\ref{Fig6} (upper figure, left side) the susceptibility as a function of the temperature only for the size $L=128$. Also in Fig.\;\ref{Fig6} (upper figure, right side) we show the log-log plot of data points of $\chi_{max}$ versus $L$ for $L= 64 , 128, 256$ and $512$ (we have excluded the smaller size $L=32$ due to finite-size effects, as observed earlier). If a first-order transition occurs in the system, the peaks $\chi_{max}$ of the susceptibility scale as
\begin{equation}\label{1_scale}
\chi_{max}\sim L^{d},
\end{equation}
where $d$ is the dimension of the lattice \cite{fisher_berker} (in our case, $d=2$). Fitting data, we have observed that $\chi_{max}$ scales as $\chi_{max}\sim L^{a}$, where
\begin{equation}\label{eq5}
a=2.087\pm 0.036,
\end{equation}
which is compatible with the form given in Eq. (\ref{1_scale}). Hence, for $\sigma=0.2$ and $h_{o}=2.0$, our simulations suggest that a first-order phase transition occurs in the system for low temperatures. We have verified that a similar behavior (jumps in the magnetization, with susceptibility diverging with $L^{2}$) also occurs for other values of the width of the distribution $\sigma$, like $\sigma=0.1$ and $0.3$ (we do not show these results because they are similar to the magnetization curves of Fig.\;\ref{Fig5}). Numerically, it is difficult to determine the threshold value $\sigma_{crit}$ for which we only have continuous phase transitions, as it was done analytically in \cite{meu}, for the equilibrium (quenched) case. However, for all of the simulations with $\sigma>0.3$ we did not detect first-order transition features.

\begin{figure}[t]
\begin{center}
\vspace{1.0cm}
\includegraphics[width=0.4\columnwidth,angle=0]{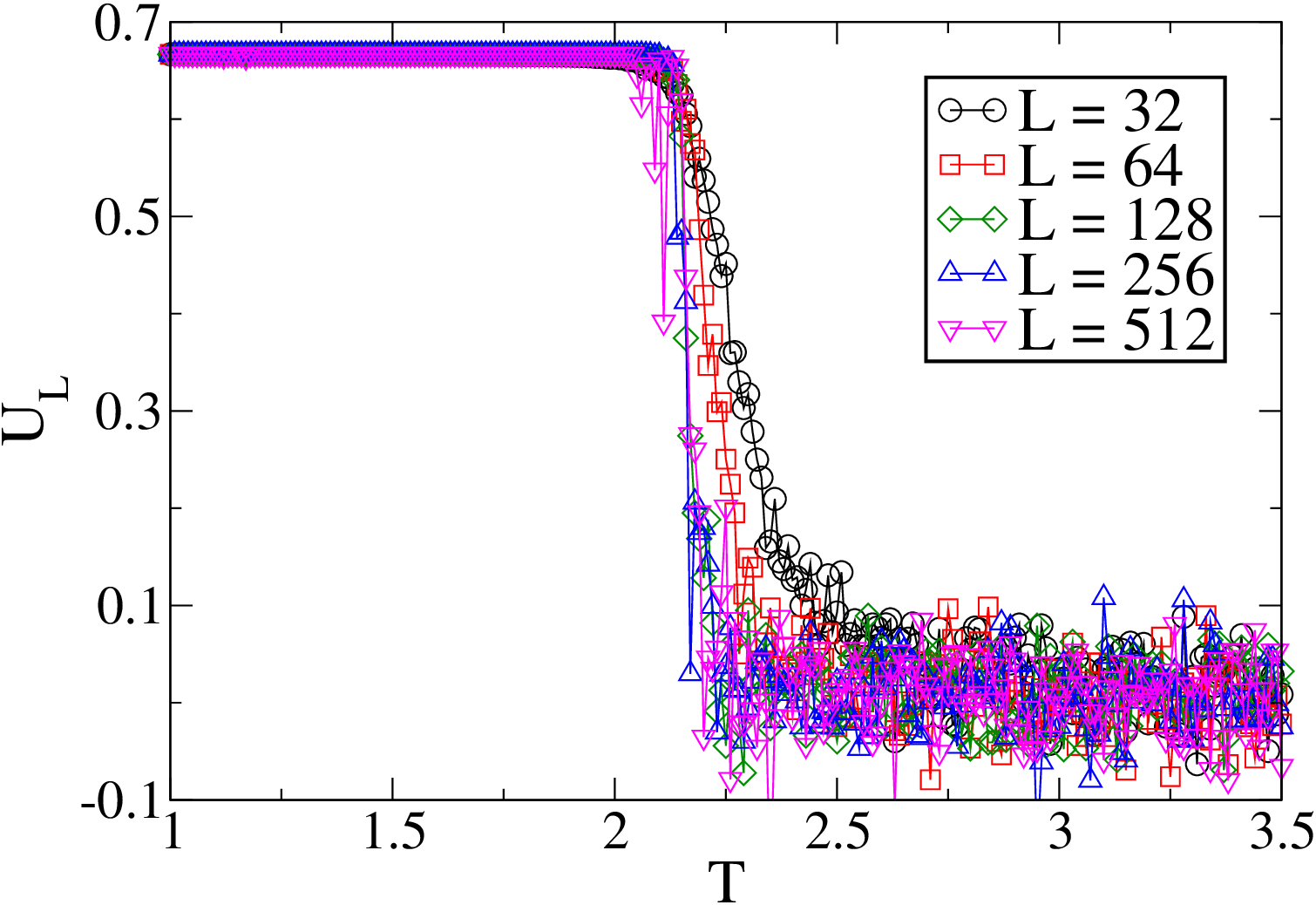}
\hspace{0.5cm}
\includegraphics[width=0.4\columnwidth,angle=0]{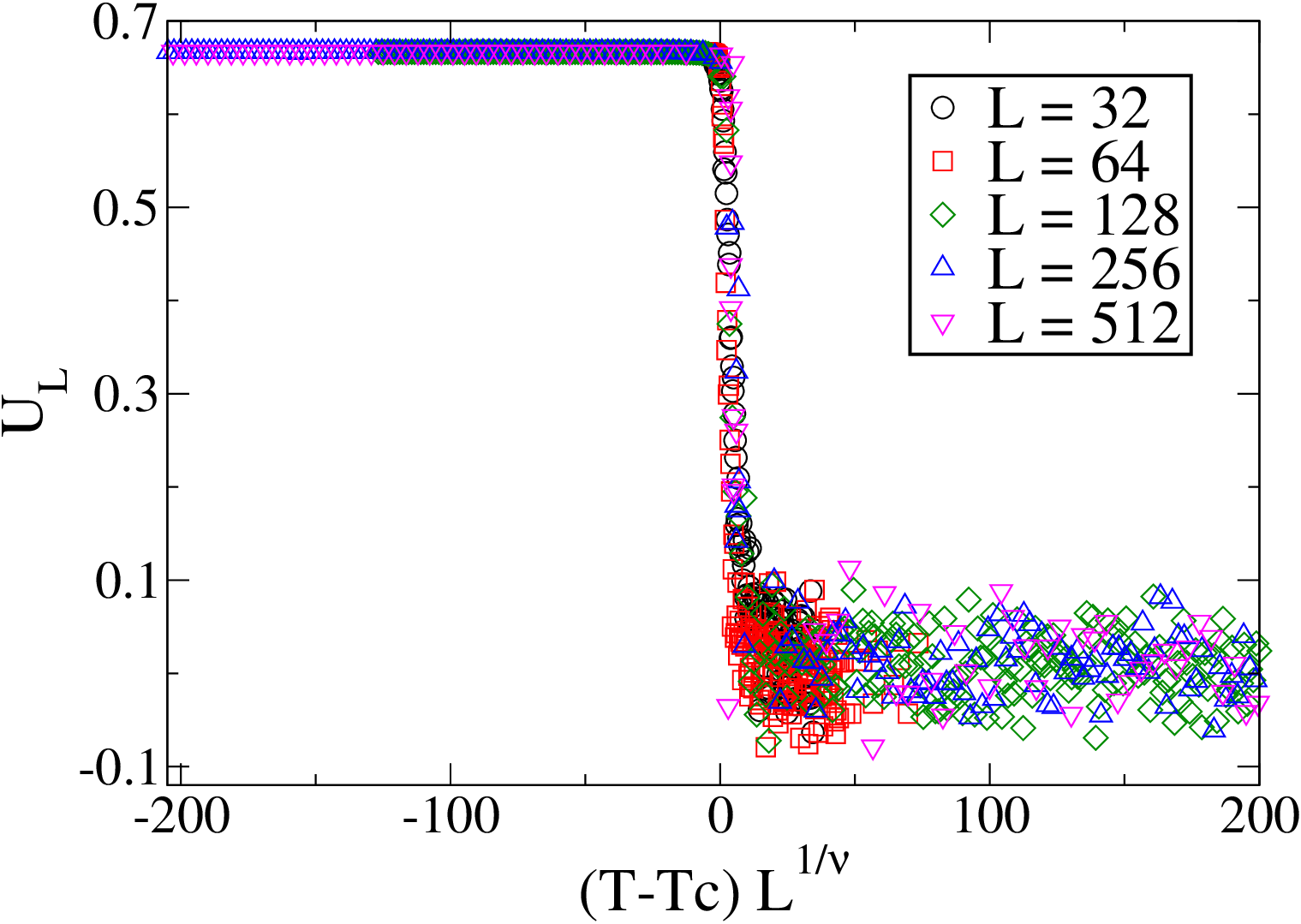}
\end{center}
\caption{Binder cumulant $U_{L}$, Eq. (\ref{eq7}), for $\sigma=0.2$, $h_{o}=0.5$ and some lattice sizes as a function of temperature (left side) and the best collapse of data (right side), based on Eq. (\ref{eq8}). The critical values are $T_{c}=2.1258$ and $\nu=1.0277$, as discussed in the text. The error bars are smaller than data points.}
\label{Fig7}
\end{figure}

\begin{figure}[t]
\begin{center}
\vspace{1.0cm}
\includegraphics[width=0.4\columnwidth,angle=0]{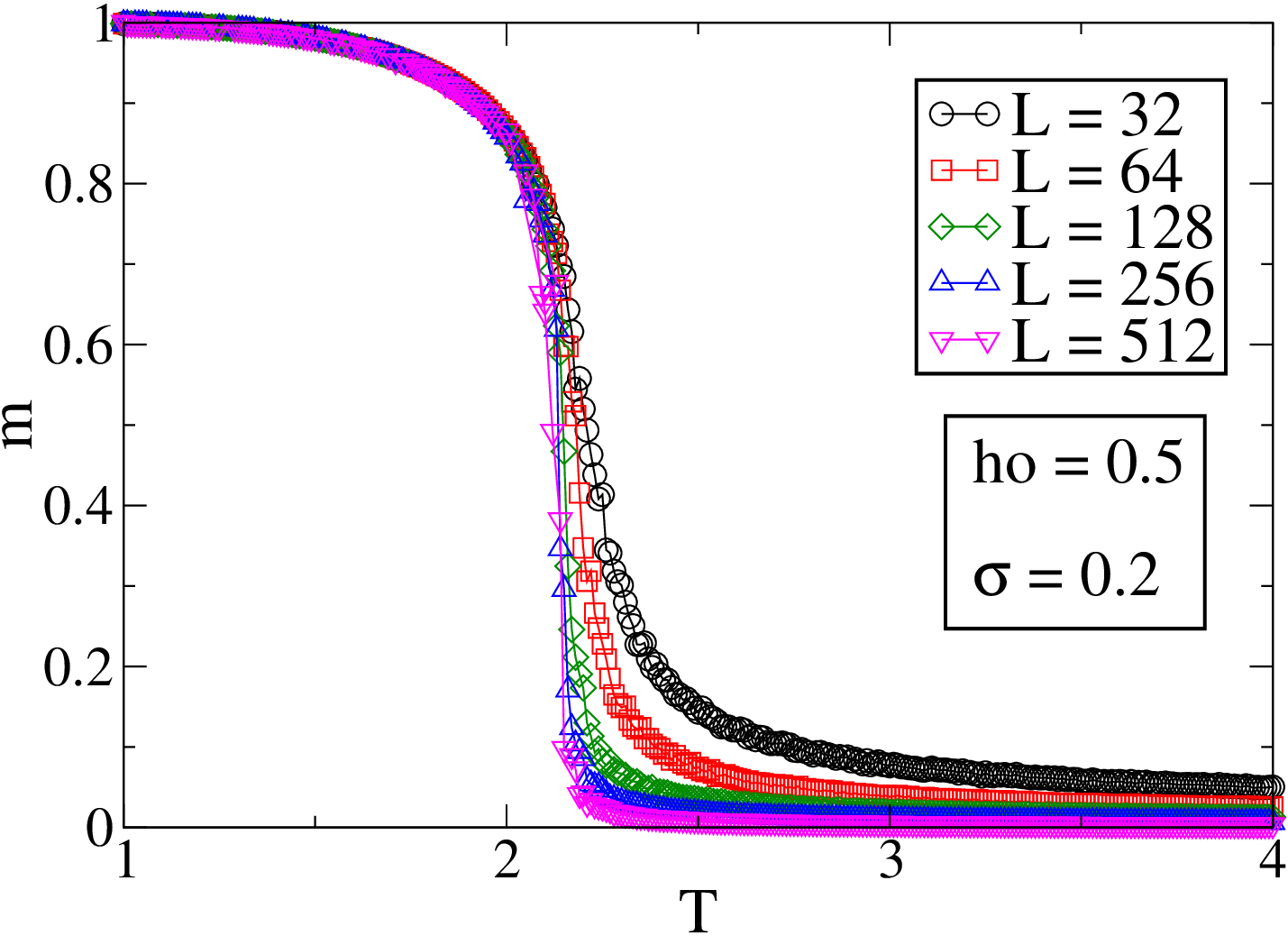}
\hspace{0.5cm}
\includegraphics[width=0.4\columnwidth,angle=0]{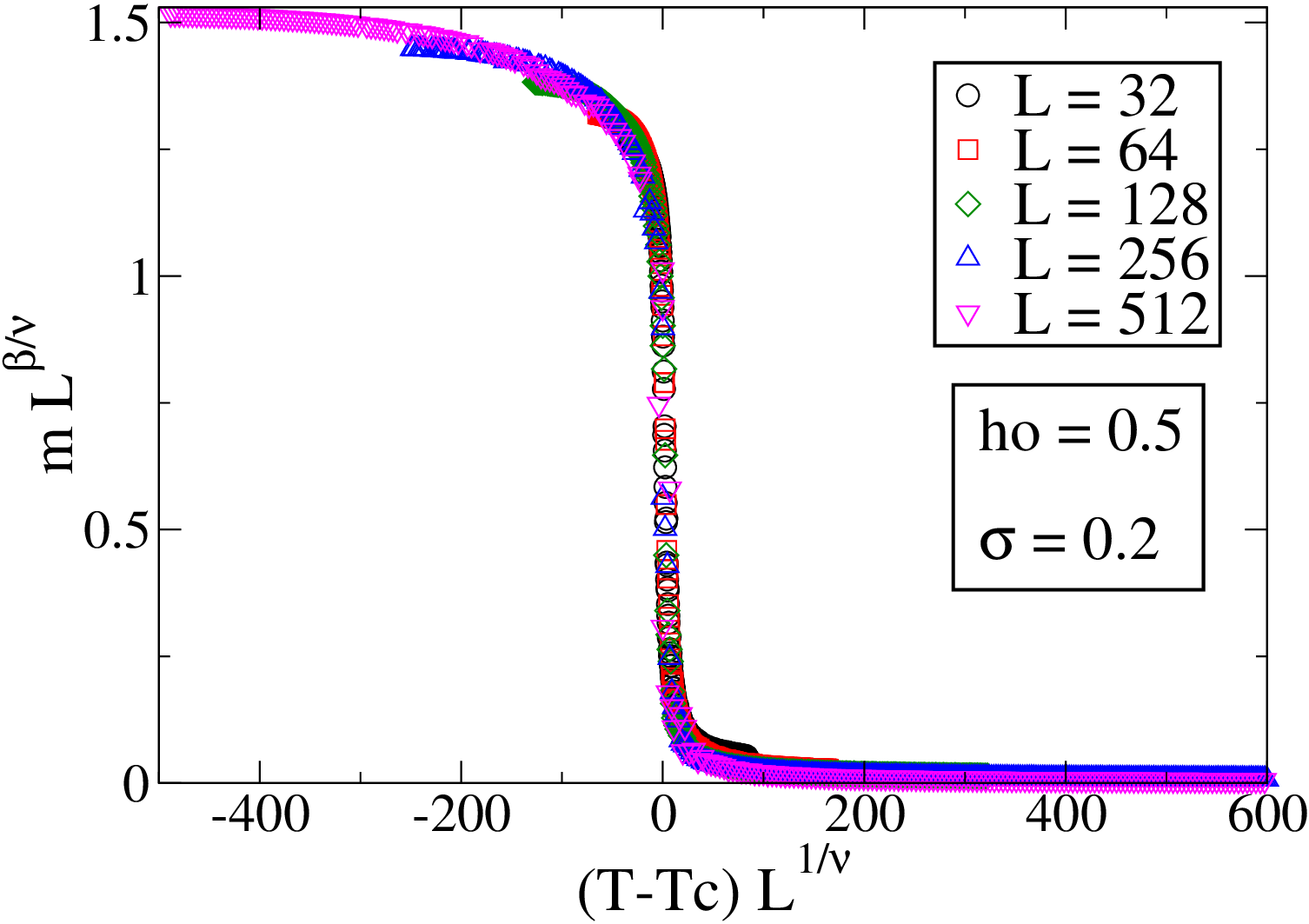}
\end{center}
\caption{Magnetization per spin as a function of temperature (left) and the scaling plot of the magnetization (right). The best collapse was obtained for $T_{c}=2.1258$, $\beta=0.0688$ and $\nu=1.0277$, as discussed in the text. The error bars are smaller than data points.}
\label{Fig8}
\end{figure}

\begin{figure}[t]
\begin{center}
\vspace{1.0cm}
\includegraphics[width=0.5\columnwidth,angle=0]{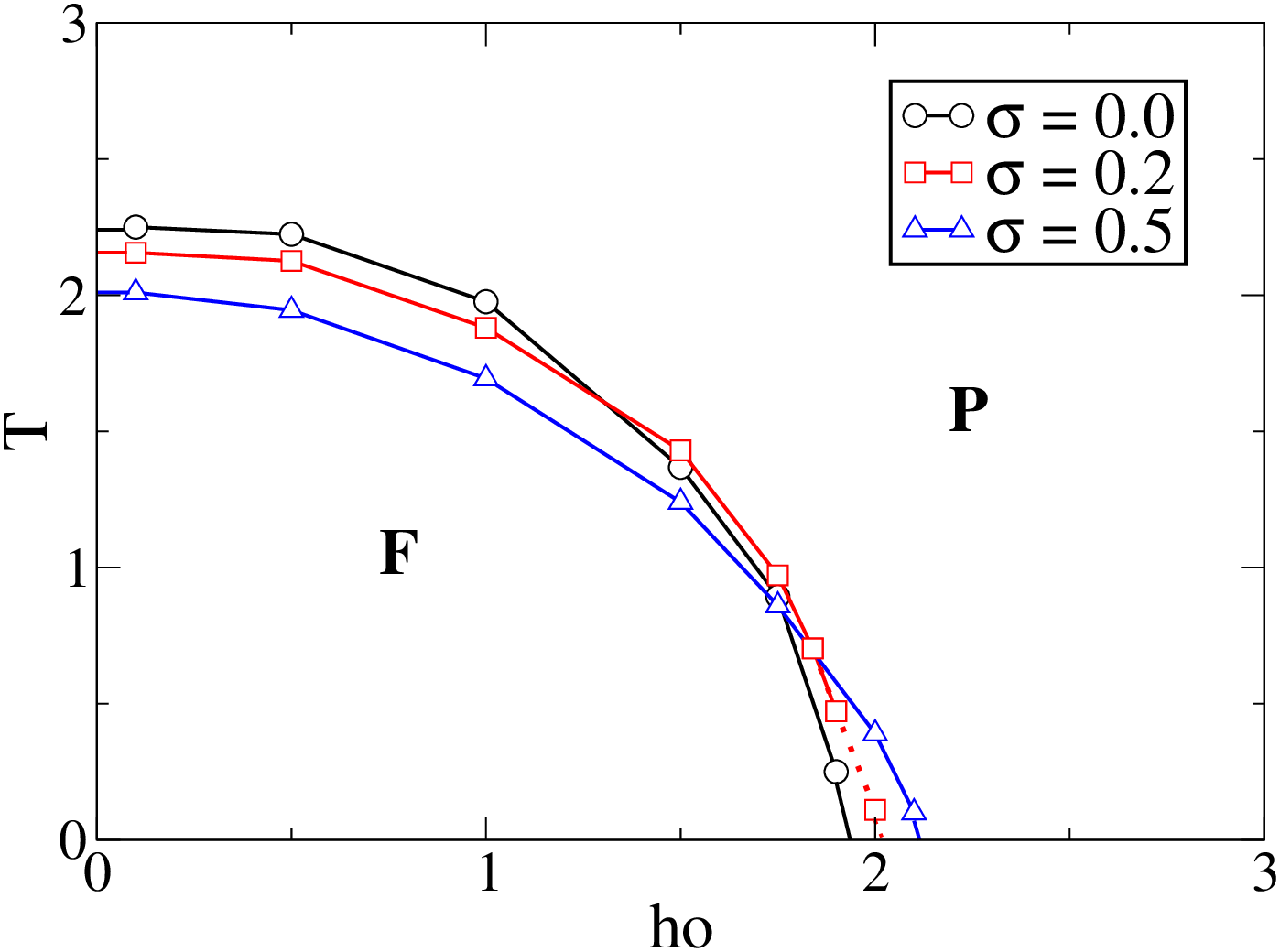}
\end{center}
\caption{Sketch of the phase diagram of the model, separating the paramagnetic (\textbf{P}) and ferromagnetic (\textbf{F}) phases, for $\sigma=0.0, 0.2$ and $0.5$. The symbols (circle, square and diamond) are computed points, whereas the lines are just guides to the eye. Full (dotted) lines represent continuous (first-order) phase transitions. As discussed in the text, we have first-order phase transitions for low temperatures and high random-field intensities only for some small values of $\sigma$. In the limit of the bimodal probability distribution ($\sigma\to0$) and for large $\sigma$, we only have continuous phase transitions. The error bars are smaller than data points.}
\label{Fig9}
\end{figure}

We have performed a preliminary calculation of the critical exponents in the low-field region in order to test whether the system follows continuous FSS laws. Thus, we have choosen the values $\sigma=0.2$ and $h_{o}=0.5$, for which our simulations suggest a continuous phase transition (see Fig.\;\ref{Fig4}), and we have calculated the magnetization and the susceptibility for various lattice sizes.
The critical temperature of the infinite lattice $T_{c}=T_{c}(\infty)$ was obtained by extrapolating the $T_{c}(L)$ values given by the susceptibility peaks positions, and for this case we have $T_{c}=2.1258\pm 0.0015$. The exponent related to the divergence of the correlation lenght $\nu$ was calculated by means of the Binder cumulant \cite{binder}, defined as
\begin{equation}\label{eq7}
U_{L}=\left[1-\frac{<m^{4}>}{3<m^{2}>^{2}}\right],
\end{equation}
\noindent
which has the FSS form
\begin{equation}\label{eq8}
U_{L}=\tilde{U_{L}}((T-T_{c})\;L^{1/\nu}),
\end{equation}
\noindent
where $\tilde{U_{L}}$ is a scaling function that is independent of $L$. In Fig.\;\ref{Fig7} we show the Binder cumulant for some lattice sizes (left side) and the best colapse of data (right side), based on Eq. (\ref{eq8}), obtained with the above-mentioned value of $T_{c}$ and $\nu=1.0277\pm 0.0102$. The exponent $\beta$, which characterizes the behavior of the magnetization near the critical point $T_{c}$, was calculated to give us the best collapse of the magnetization curves. In Fig.\;\ref{Fig8} we show the magnetization per spin versus the temperature (left side) and the best collapse of the curves (right side), based on the standard FSS forms
\begin{eqnarray}\nonumber
T_{c}(L) & = & T_{c}-a\;L^{-1/\nu}, \\ \label{eq6}
m(T,L) & = & L^{-\beta/\nu}F((T-T_{c})\;L^{1/\nu}),
\end{eqnarray}
\noindent
where $a$ is a constant and $F(x)$ is a scaling function that has the limiting behaviors $F(x<<1)\sim x^{\beta}$ and $F(x>>1)=c$, where $c$ is a constant. From the above FSS forms, Eqs. (\ref{eq6}), we have found that $\beta=0.0688\pm 0.0032$. We have verified again that the susceptibilty peak grows very fast, and we show in Fig.\;\ref{Fig6} (lower figure, left side) the susceptibility as a function of the temperature only for $L=128$, in order to compare the shape of this curve with the one obtained for a first-order transition (Fig.\;\ref{Fig6}, upper figure, left side). Also in Fig.\;\ref{Fig6} (lower figure, right side), we show the peaks of the susceptibility versus the linear lattice size $L$ in the log-log scale for $L=32, 64, 128, 256$ and $512$. For the continuous transition, the finite size effects are not so pronounced as in the first-order case, and we have considered even the smaller lattice size, $L=32$. These peaks scale with the system size in the traditional power-law form for continuous transitions, i.e.,

\begin{eqnarray}\nonumber
\chi_{max}\sim L^{\gamma/\nu},
\end{eqnarray}
\noindent
where $\gamma/\nu=1.7157\pm0.0499$. For the value of $\nu$ obtained above, we have $\gamma=1.7632\pm 0.0005$. In other words, the standard FSS forms for continuous transitions, Eqs. (\ref{eq8}) and (\ref{eq6}), are the correct FSS equations for that nonequilibrium model. It is the purpose of this exponents' calculation; how the critical exponents depend on the choice of the parameters of the disorder distribution ($\sigma$ and $h_{o}$) is a question beyond the target of this work. To summarize the results discussed along this paper, we show in Fig. \ref{Fig9} a sketch of the phase diagram of the model, in the plane temperature $T$ versus random field intensity $h_{o}$ for some values of the parameter $\sigma$.

\section{Conclusions}
In this work, we have studied the nonequilibirum random-field Ising model on a square lattice with nearest-neighbors interactions by means of Monte Carlo simulations. The lattice sizes analyzed were $L=32, 64, 128, 256$ and $512$. Due to computational cost, we have used only $10^{6}$ MC steps per spin for averaging. In order to generate the random fields, we have used the recently proposal double-gaussian probability distribution, which consists of two single gaussian distributions, centered at $+h_{o}$ and $-h_{o}$, with the same width $\sigma$. As pointed in a previous work \cite{meu}, this distribution is expected to be more appropriate for the theoretical description of real systems than other simpler well-known cases, i.e., the bimodal and single gaussian distributions.

The time evolution of the system is stochastic because of a competing spin-flip kinetics, which, in addition to the usual heat bath, involves a random external magnetic field. The competition induces a kind of dynamical disorder that might be present in real disordered systems such as the class of random-field materials \cite{lacomba2}. This system differs from the standard equilibrium ones: while the local field is randomly assigned in space according to a distribution $P(h_{i})$, which remains frozen in for the quenched case, and $P(h_{i})$ contains essential correlations in the annealed system, where the impurity distribution is in equilibrium with the spin system, our case is similar to the quenched system at each time during the stationary regime, but $h_{i}$ keeps randomly changing with time, also according to $P(h_{i})$, at each site $i$.

Our results suggest that in the limit of the bimodal distribution ($\sigma=0.0$), the transition from the ordered to the disordered phase is continuous. However, first-order phase transitions may occur in the system for low temperatures and high random-field intensities $h_{o}$, for some values of the parameter $\sigma$. The order of the transition was determined by the scaling of the susceptibility peaks, which must grow with the total number of spins $L^{2}$ in the case of first-order transitions \cite{fisher_berker}. The threshold value of $\sigma$, for which the phase transitions are always continuous, is difficult to determine numerically. Nonetheless, the simulations suggest that for $\sigma> 0.3$ first-order transitions do not occur anymore.

We have performed a preliminary calculation of the critical exponents in the low-field region in order to test whether the system follows continuous FSS laws. So, we have choosen the values $\sigma=0.2$ and $h_{o}=0.5$, for which our simulations suggest a continuous phase transition, and we have calculated the critical exponents. We have found that the system obeys standard continuous FSS laws, with $\beta=0.0688\pm 0.0032$, $\nu=1.0277\pm 0.0102$ and $\gamma=1.7632\pm 0.0005$. Note that the susceptibility peaks grow with the system size $L$ with an exponent $\gamma/\nu < 2$, as expected for continuous phase transitions in two-dimensional lattices.

The mean field analysis of the Random Field Ising Spin Glass (RFISG) that follows the double-gaussian probability distribution was done recently for the case of quenched variables \cite{meu2}. That study showed a rich behavior of the system, with continuous and first-order phase transitions as well as a change in the concavity of the Almeida-Thouless (AT) line \cite{at}, which was experimentally verified in the diluted antiferromagnets $Fe_{x}Zn_{1-x}F_{2}$ \cite{montenegro}. However, recent numerical results suggest that there is no AT line for short-range models with bond disorder and/or under random magnetic fields \cite{young1,sasaki,jorg}. Thus, extensions of this work concerning on MC simulations of the RFISG in the presence of random fields generated according to the double-gaussian distribution, for the equilibrium (quenched and annealed variables) and nonequilibrium cases, would be of great interest to elucidate the existence of the AT line in equilibrium systems with short-range interactions and to obtain more results for nonequilibrium spin glasses.

\section*{Acknowledgments}
The author thanks O. R. Salmon for fruitful discussions and D. O. Soares-Pinto and Silvio M. Duarte Queir\'os for their comments. Financial support from the Brazilian funding agency CNPq is also acknowledge.


\section*{References}
\begin{thebibliography}{30}

\bibitem{binder_review}
K. Binder, A. P. Young, Rev. Mod. Phys. \textbf{58}, 801 (1986).

\bibitem{belanger_review}
D. P. Belanger, T. Nattermann, {\it Spin Glasses and Random Fields}, edited by A.~P. Young (World Scientific, Singapore, 1998).

\bibitem{dotsenko}
V. S. Dotsenko, J. Stat. Mech. \textbf{P09005} (2007).

\bibitem{fishman}
S. Fishman, A. Aharony, J. Phys. C: Solid State Phys. \textbf{12}, L729 (1979).

\bibitem{cardy}
J. Cardy, Phys. Rev. B \textbf{29}, 505 (1984).

\bibitem{belanger}
D. P. Belanger, A. R. King, V. Jaccarino, J. L. Cardy, Phys. Rev. B \textbf{28}, 2522 (1983).

\bibitem{birgeneau}
R. J. Birgeneau, J. Magn. Magn. Mater. \textbf{177}, 1 (1998).

\bibitem{dickman}
J. Marro and R. Dickman, {\it Nonequilibrium Phase Transitions in Lattice Models} (Cambridge University Press, Cambridge, 1999).

\bibitem{meu}
N. Crokidakis and F.~D. Nobre, J. Phys.: Condens. Matter \textbf{20}, 145211 (2008).

\bibitem{thorpe}
M. F. Thorpe and D. Beeman, Phys. Rev. B \textbf{14}, 188 (1976).

\bibitem{urumov}
V. Urumov, J. Phys.: Condens. Matter \textbf{1}, 7037 (1989).

\bibitem{haroni}
A. A. Haroni and C. E. Paraskevaidis, Phys. Stat. Sol. (b) \textbf{193}, 445 (1996).

\bibitem{goncalves}
L. L. Gon\c{c}alves and R. B. Stinchcombe, Phys. Rev. B \textbf{33}, 4762 (1986).

\bibitem{vieira}
A. P. Vieira and L. L. Gon\c{c}alves, Journ. Chem. Phys. \textbf{110}, 1235 (1999).

\bibitem{lacomba1}
A. I. L\'opez-Lacomba and J. Marro, Europhys. Lett. \textbf{25}, 169 (1994).

\bibitem{lacomba2}
A. I. L\'opez-Lacomba and J. Marro, Phys. Rev. B \textbf{46}, 8244 (1992).

\bibitem{garrido}
P. L. Garrido and J. Marro, Europhys. Lett. \textbf{15}, 375 (1991).

\bibitem{miranda}
J. M. Gonz\'alez-Miranda, A. Labarta, M. Puma, J. F. Fern\'andez, P. L. Garrido and J. Marro, Phys. Rev. E \textbf{49}, 2041 (1994).

\bibitem{bonilla}
L. L. Bonilla, F. G. Padilla, G. Parisi and F. Rittort, Phys. Rev. B \textbf{54}, 4170 (1996).

\bibitem{metropolis}
N. Metropolis, A. W. Rosenbluth, N. M. Rosenbluth, A. H. Teller and E. Teller, J. Chem. Phys. \textbf{21}, 1087 (1953).

\bibitem{aharony}
A. Aharony, Phys. Rev. B \textbf{18}, 3318 (1978).

\bibitem{schneider}
T. Schneider and E. Pytte, Phys. Rev. B \textbf{15}, 1519 (1977).

\bibitem{andelman}
D. Andelman, Phys. Rev. B \textbf{27}, 3079 (1983).

\bibitem{gofman}
M. Gofman, J. Adler, A. Aharony, A. B. Haris and M. Schwartz, Phys. Rev. Lett. \textbf{71}, 1569 (1993); Phys. Rev. B {\bf 53}, 6362 (1996).

\bibitem{swift}
M. R. Swift, A. J. Bray, A. Maritan, M. Cieplak and J. R. Banavar, Europhys. Lett. \textbf{38}, 273 (1997).

\bibitem{gould}
H. Gould and J. Tobochnik, {\it An Introduction to Computer Simulation Methods: Applications to Physical Systems} (Addison-Wesley Publishing Company, Reading MA, 1996).

\bibitem{fisher_berker}
M. E. Fisher and A. N. Berker, Phys. Rev. B \textbf{26}, 2507 (1982).

\bibitem{binder}
K. Binder, Z. Phys. B \textbf{43}, 119 (1981).

\bibitem{meu2}
N. Crokidakis and F. D. Nobre, Phys. Rev. E \textbf{77}, 041124 (2008).

\bibitem{at}
J. R. L. de Almeida and D. J. Thouless, J. Phys. A \textbf{11}, 983 (1978).

\bibitem{montenegro}
F. C. Montenegro, A. R. King, V. Jaccarino, S. -J. Han and D. P. Belanger, Phys. Rev. B \textbf{44}, 2155 (1991).

\bibitem{young1}
A. P. Young and H. G. Katzgraber, Phys. Rev. Lett. \textbf{93}, 207203 (2004).

\bibitem{sasaki}
M. Sasaki, K. Hukushima, H. Yoshino and H. Takayama, Phys. Rev. Lett. \textbf{99}, 137202 (2007).

\bibitem{jorg}
T. Jorg, H. G. Katzgraber and F. Krzakala, Phys. Rev. Lett. \textbf{100}, 197202 (2008).

\end{thebibliography}
\end{document}